\documentstyle[12pt,epsfig]{article}
\def\hor{\hskip 0.8cm}

\begin{document}
%17 november 1998.

\rightline{LYCEN 98102}

\begin{center}
{\Large \bf The proton structure function and \\
            a soft Regge Dipole Pomeron~: \\
            a test with recent data}

\bigskip
\bigskip

{\bf P. Desgrolard}({\footnote{E-mail: desgrolard@ipnl.in2p3.fr}}),
{\bf A. Lengyel}({\footnote{E-mail: sasha@len.uzhgorod.ua}}),
{\bf E. Martynov}({\footnote{E-mail: martynov@bitp.kiev.ua}}).

\end{center}

\bigskip

($^{1}$){\it 
Institut de Physique Nucl\'eaire de Lyon,\\ 
IN2P3-CNRS et Universit\'{e} Claude Bernard,\\
43 boulevard du 11 novembre 1918, F-69622 Villeurbanne Cedex, France\\}
($^{2}$){\it Institute of Electronic Physics,\\
National Academy of Sciences of Ukraine,\\
294015 Uzhgorod-015, Universitetska 21, Ukraine\\}
($^3$){\it
N.N. Bogoliubov Institute for Theoretical Physics,\\
National Academy of Sciences of Ukraine,\\
252143, Kiev-143, Metrologicheskaja 14b, Ukraine\\}

\bigskip
\bigskip

{\bf Abstract}
A recently published soft Regge Dipole Pomeron model intended for all $x$ and
$Q^2$ is proved to give a good
agreement with (non fitted) recent HERA data from ZEUS (SVX95) on the proton
structure function $F_2^p(x,Q^2)$ at low $Q^2$ and low $x$. 
The model also reproduces (without fit) the recently estimated 
experimental derivatives ${\partial F_2^p\over\partial\ell n Q^2}$ and 
${\partial \ell n F_2^p\over\partial\ell n (1/x)}$ in a wide $x$ and 
$Q^2$-region.

\bigskip
\bigskip
\bigskip

{\bf 1 Motivations}
\medskip

The proton structure function (SF) is one of the observables most often measured
in high energy physics~\cite{ICHEP98}. Consequently a relevant model has to be 
periodically tested (and eventually updated or abandoned) in the new 
kinematical ranges of $x$ (Bj\"orken variable) and $Q^2$ (virtuality of the
photon) investigated by the experimentalists. 

\hor
Recent measurements~\cite{ZEUS98} of the SF at HERA 
(from ZEUS 1995 shifted vertex experiment (SVX95)) have motivated us to 
test our Dipole Pomeron parametrization~\cite{dlm} of the proton structure
function $F_2^p(x,Q^2)$ intended for a wide region of $Q^2$ and $x$. 
We wish to show that these
recent data can be reproduced within an "old soft Pomeron" framework, which
is an "\`a la Regge" approach. 

\hor
Second, we revise a widely extended opinion that a soft Pomeron and more
generally an "\`a la Regge" approach to deep inelastic scattering (DIS)
should be restricted to
rather small values of $Q^2$. This conclusion is based mainly on the popular
Donnachie-Landshoff (DL) model of the 
Pomeron~\cite{dl} and its particular parametrization of $Q^2$-dependence of the
residue function\cite{dl1}.  This model was used in~\cite{ZEUS98} and the
conclusion was drawn that the Regge theory describes well the data only at
very low $Q^2\leq 0.65$ GeV$^2$. We have {\it vice-versa} shown that a 
soft Pomeron contribution (with unit intercept) can be applied to
the virtual photoproduction cross-section~\cite{dlm},
and that it reproduces well the data on $F_2^p(x,Q^2)$ in 
a much wider region of $Q^2$ and $x$.  

\hor
In the present short note, we emphasize that the fit~\cite{dlm} not only
reproduces with high quality also the new data on the SF~\cite{ZEUS98} at low 
$Q^2$ and low $x$, but is in good agreement with the behavior 
of the derivatives
${\partial F_2^p\over\partial\ell n Q^2} $
and 
${\partial \ell n F_2^p\over\partial\ell n (1/x)}$
, measured in~\cite{ZEUS98},
up to intermediate values of the kinematical variables. 

\bigskip
{\bf 2 The model}
\medskip

Many Pomeron models are on the market in high energy hadron phenomenology. In 
spite of the quite different $t$-dependence of the elastic amplitudes,
at $t=0$ they can be combined in two groups:
\begin{enumerate}
\item
A simple pole in the complex angular momenta ($j$-) plane with an intercept
$\alpha_{\cal P}(0)>1$~:  the DL Pomeron~\cite{dl} and its generalization
\cite{dglm} with an additional constant term 
(a preasymptotic simple $j$-pole with unit intercept) are examples.
Such a Pomeron leads to a total cross-section 
$$
\sigma_{tot}(s)\propto s^{\alpha_{\cal P}(0)-1} \ ,
$$
which at extremely high energies ( well beyond the present attainable ones)
would violate the Froissart-Martin bound; this could require 
to be unitarized (for example, by an eikonal method).
\item 
More complicate singularity in the $j$-plane at $j=1$~: in these 
models (\cite{dlm,dglm,jenk,nic} and references therein)
the Froissart-Martin bound is not violated and asymptotically the
total cross-sections behave as
$$
\sigma_{tot}(s)\propto \ell n^\mu (s/s_0), \quad 0<\mu < 2, 
\quad s_0=1 \ {\rm GeV}^2.
$$
\end{enumerate}

All these models describe quite well hadronic total cross-sections
and $\gamma p$ inelastic one. The best quality of the 
description (in the sense of $\chi^2$) was achieved~\cite{dglm} at $t=0$ when 
$\mu =1$.
This corresponds to a double $j$-pole in the forward amplitude, {\it i.e.}
to the Dipole Pomeron (DP) model. This model was succesfully applied~\cite{dlm} 
(with its extension to $Q^2\neq 0$) to DIS with a good description of the SF
in a wide region of $Q^2$ and $x$.

\hor
Defining the DP model for DIS, we start from the expression
connecting the transverse cross-section
%$\sigma_T^{\gamma^* p}$ 
for the ($\gamma^* p$) interaction to the proton 
structure function $F_2^p$
\begin{equation}
\sigma_T^{\gamma^* p}(W,Q^2)=\frac{4\pi^2\alpha 
}{Q^2}(1+\frac{4m_p^2x^2}{Q^2})\frac{1}{1+R(x,Q^2)}F_2^p(x,Q^2)\ ,
\end{equation}
where $\alpha$ is the fine structure constant, $m_p$ is the proton mass, $R$ is 
the ratio of longitudinal to transverse cross-sections. We have approximated 
$R(x,Q^2)=0$, due to its experimental smallness. The center of mass energy 
$W$ of the ($\gamma^* p$) system obeys
\begin{equation}
W^2=Q^2 \left({1\over x}-1\right)+m^2_p 
\end{equation}
and the optical theorem writes
\begin{equation}
F_2^p(x,Q^2)=\frac{Q^2(1-x)}{4\pi^2\alpha (1+4m_p^2x^2/Q^2)}\ 
\Im {\rm m} A(W^2,t=0,Q^2) \ .
\end{equation}
The forward $\gamma^* p$ scattering amplitude is dominated, for $W$ far 
from the threshold $W_{th}=m_p$, by the Pomeron and $f$-Reggeon contributions
(we ignore an $a_2$-Reggeon contribution considering the $f$-term as an
effective one at $W> 3$ GeV)
\begin{equation}
A(W^2,t=0,Q^2)=P(W^2,Q^2)+F(W^2,Q^2)\ ,
\end{equation}
with
\begin{equation}
F(W^2,Q^2)=iG_f(Q^2)\bigg (-i\frac{W^2}{m_p^2}\bigg 
)^{\alpha_f(0)-1}(1-x)^{B_f(Q^2)}
\end{equation}
and
\begin{equation}
P(W^2,Q^2)=P_1+P_2 \ ,
\end{equation}
\begin{equation}
P_1=iG_1(Q^2)\ell n(-iW^2/m_p^2)(1-x)^{B_1(Q^2)}\ , \qquad 
P_2=iG_2(Q^2)(1-x)^{B_2(Q^2)}\ .
\end{equation}
The details of the parametrization of the real functions $G_i(Q^2),\,B_i(Q^2)$ 
can be found in~\cite{dlm}. Here we only mention that they vary between the 
constants $G_i(0),\,B_i(0)$ and $G_i(\infty),\, B_i(\infty)$.

\hor
Let us  make a few comments on the chosen Pomeron model.

We support the  point of view that there is just one "bare" Pomeron (two 
Pomerons, "soft" and "hard", are considered in~\cite{dl2}). This unique Pomeron
is universal and 
factorizable\footnote{If Pomeron is a sum of two terms (as in 
(7)) then at least the leading one at $W\gg m_p$ should satisfy  
factorization.},
{\it  i.e.} it is the same in all processes; only the vertex 
functions depend on which are the interacting particles. It follows from 
these special requirements that the Pomeron trajectory should be independent 
of $Q^2$. 
One should mention that this approach differs from other
ones where an "effective" Pomeron with a 
$Q^2$-dependent intercept~\cite{ckmt,bgp,al,djp}) has been chosen.
The hard Pomeron or BFKL Pomeron~\cite{bfkl} with a quite large intercept is 
only an approximation to a true Pomeron. A growth of the total 
cross-sections means that in $j$-plane a true Pomeron is harder than a 
simple pole singularity at $j=1$.

The Dipole Pomeron defined by (6),(7) obeys the following specificities: 
it is universal, it has a $Q^2$-independent intercept 
$\alpha_P(0)=1$, it does not violate, at least explicitly, the unitarity 
restrictions on the amplitude.   

\bigskip
{\bf 3 The proton SF and total ($\gamma p$) cross section results}
\medskip

We choose the DP model defined in details in~\cite{dlm} with the 23
parameters (see Table~2 in~\cite{dlm}).  This is not probably the most 
economical set
within this framework, however we keep these published values
and do not perform any new adjustement in order to
introduce no confusion (we postpone an update of this model, fixing some
parameters and refitting the others with a set of data completed by the
forthcoming 96-97 HERA results).  
We have proved that adding the new (44) values
of $F_2^p$ does not change the quality of the fit in~\cite{dlm}, we found~:
$\chi^2=1321$ for 1209 data points which becomes $\chi^2=1341$ for 1253 data, 
in practice leaving unchanged $\chi^2_{/d.o.f}\simeq 1.1$.

\hor
The $F_2^p(x,Q^2)$ results are plotted versus $x$ for the experimental $Q^2$ 
bins of~\cite{ZEUS98} (low $Q^2$) and compared to the whole set of fitted 
and non fitted data in Fig.~1. 
The total real ($\gamma p$) cross section versus the c.m squared energy $W^2$ is 
shown in Fig.~2.
These figures show the good agreement for $0\leq Q^2\leq 17$ GeV$^2$; higher
$Q^2$ values (where DP model also reproduce well the
data) are discussed in~\cite{dlm}.

\newpage

\bigskip
{\bf 4 The $Q$-slope as a function of $x$}
\medskip

The ${Q}$-slope 
$$B_Q(x,Q^2)={\partial F_2^p(x,Q^2)\over\partial\ell n (Q^2)}\                $$

depends on the two independent variables $x$ and $Q^2$.
However to compare with experiment the $Q$-slope has been calculated for the set

\cite{ZEUS98} 
$(x_i,\ {<Q^2>}_i)$; i=1,24
of strongly correlated variables which includes $x$ up to 0.2.                                        
The results are shown in Fig.~3; the agreement is good up to $x\sim 0.1$.

\bigskip
{\bf 5 The $x$-slope and the "effective intercept" as functions of $Q^2$}
\medskip

The $x$-slope 
$$B_x(x,Q^2)={\partial \ell n F_2^p(x,Q^2)\over\partial\ell n (1/x)}$$
is also a function of two variables;
the quantity currently replacing $B_x$ in experimental papers is the 
"effective power" $\Delta_{eff}$ (sometimes denoted as $\lambda_{eff}$) 
in the low $x$ - fixed $Q^2$ approximation  of the structure function
$$
F_2^p \propto x^{-\Delta_{eff}}.
$$
\hor
This power is currently connected to
the Pomeron effective intercept ($\alpha(0)=1+\Delta_{eff}$). Actually, from a 
phenomenological point of 
view, one can extract $\Delta_{eff}$ at fixed 
$Q^2$ depending on $x$ assuming a parametrization 
$$ F_2^p(x,Q^2)=G(Q^2)\left({1\over x}\right)^{\Delta_{eff}(x,Q^2)}\ .       $$

Strictly speaking, however, the identification 
$B_x=\Delta_{eff}$ is possible only when 
the $x-$indepen\-dence of the $x$-slope is a model assumption (one may 
see~\cite{dlm} for a discussion of the slopes). In general, this is 
not the case. From that point of view, it would be interesting 
and important to have 
"measured" values for $\Delta_{eff}$ at fixed $Q^2$ an at different $<x>$~:
it will allow to study the $x$-dependence of the effective Pomeron intercept. 

\hor
The comparison between the calculated value of the $x$-slope and the
experimental effective power $\lambda_{eff}$ is given in the last figure 
(Fig.~4) for the set of kinematical variables~\cite{ZEUS98}   
($<x>_i$, $Q_i^2);\ {i=1,30}$) including ZEUS and fixed target E665 results. 
The agreement is good in the whole experimental 
range~: up to $Q^2\sim 250$ GeV$^2$.

\bigskip
{\bf 6 Conclusion}
\medskip

We have proved that a soft dipole Pomeron model~\cite{dlm} 
("{\it \`a la }" Regge) not only reproduces with a high quality the new data 
on the proton structure function~\cite{ZEUS98} at low $Q^2$
and low $x$, but also is in good agreement with the  
measured slopes~\cite{ZEUS98} ${\partial F_2^p\over\partial\ell n Q^2} $ 
and ${\partial \ell n F_2^p\over\partial\ell n (1/x)}$
up to $Q^2\sim 250$ GeV$^2$ and $x\sim 0.1$.

\hor
Such a success in reproducing the data is due not only to an appropriate 
choice of the asymptotic Pomeron contribution but is due also to 
the preasymptotic terms (chosen constant here) in the Pomeron and 
$f$-reggeon. A more detailed discussion of preasymptotics properties of the 
model is in \cite{dlm}.

\hor
This enforces our belief that a universal -factorizable- Pomeron with a $Q^2$
independent intercept $\alpha(0)=1$ may be successful in Deep Inelastic 
Scattering 
not only at low $Q^2$ and $x$ (as the DL Pomeron does).
On the basis of the results from \cite{dlm} and the present paper, we claim that 
the area of validity of a Regge approach is 
much wider (especially in $Q^2$) than usually assumed and can be
extended up to rather high values of $Q^2$ (may be up to a few hundreds GeV$^2$). 

\medskip
{\bf Acknowledgments} E.M. would like to thank the IPNL for the kind 
hospitality and financial support provided to him during this work.
It is a pleasure to thank E. Predazzi for a critical reading of the
manuscript and illuminating comments.

\bigskip
\bigskip

%\newpage

\begin{center}
\epsfig{figure=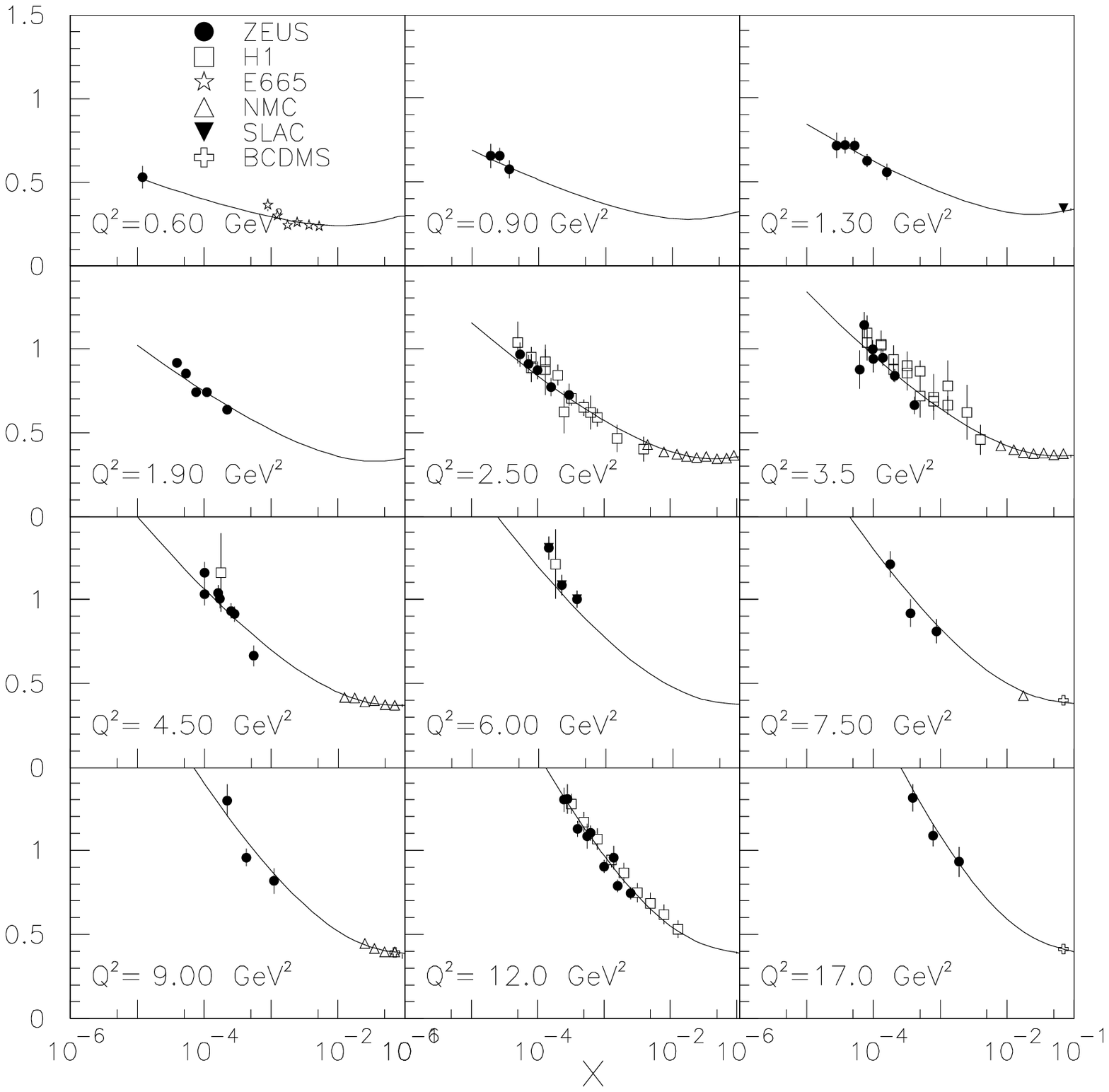,width=16cm}
\end{center}
\bigskip
{\bf Fig.~1}
Experimental data for the proton structure function
$F_2^p(x,Q^2)$ at low $Q^2$  compared to the results within the 
Dipole Pomeron model (shown are the 44 recent ZEUS SVX 95 data -non fitted-
and the other fitted data).  

\begin{center}
\epsfig{figure=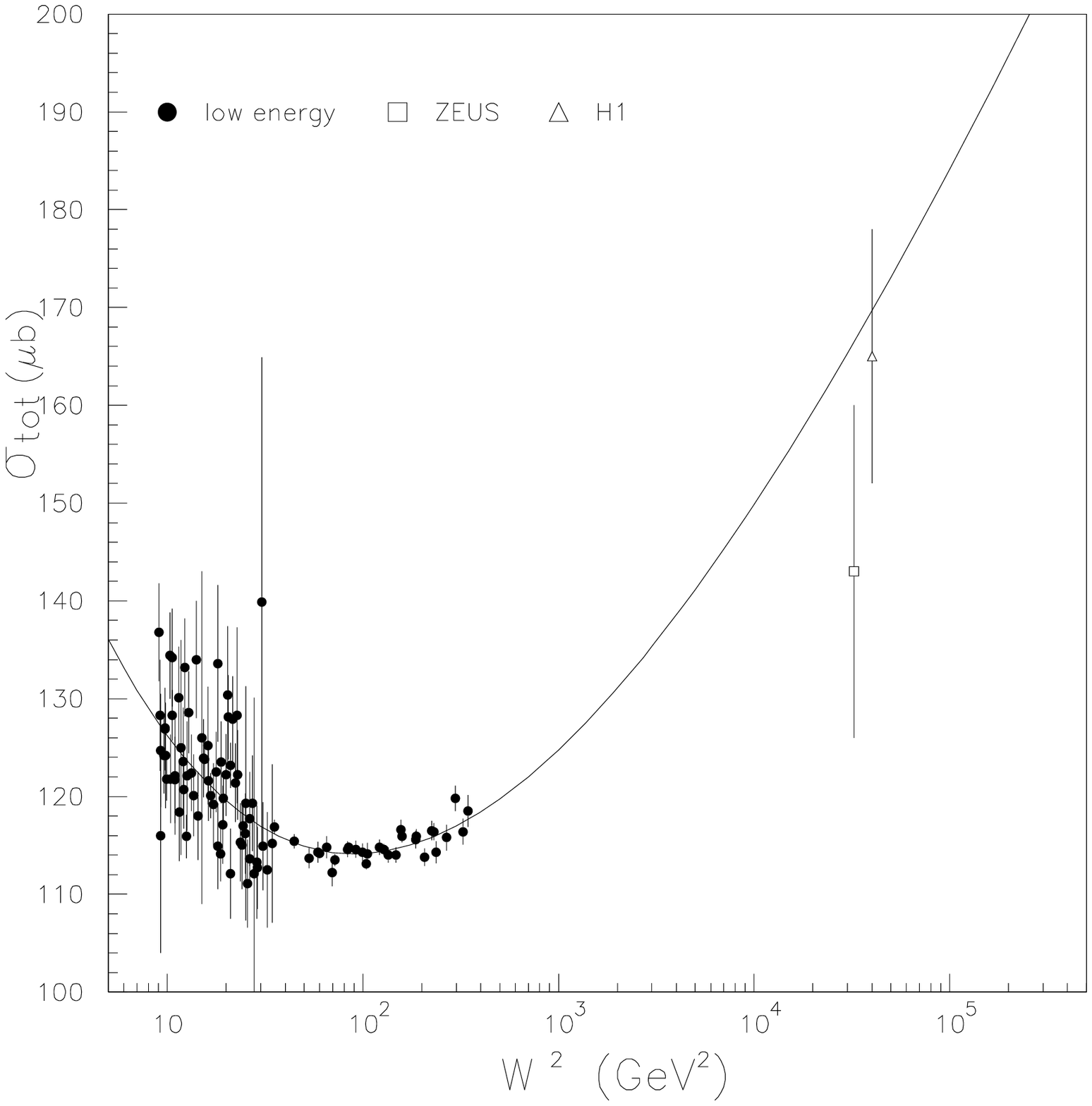,width=12cm}
\end{center}
\bigskip
{\bf Fig.~2.}
Experimental fitted data for the photoproduction total cross-section
and predictions in the Dipole Pomeron model.

\begin{center}
\epsfig{figure=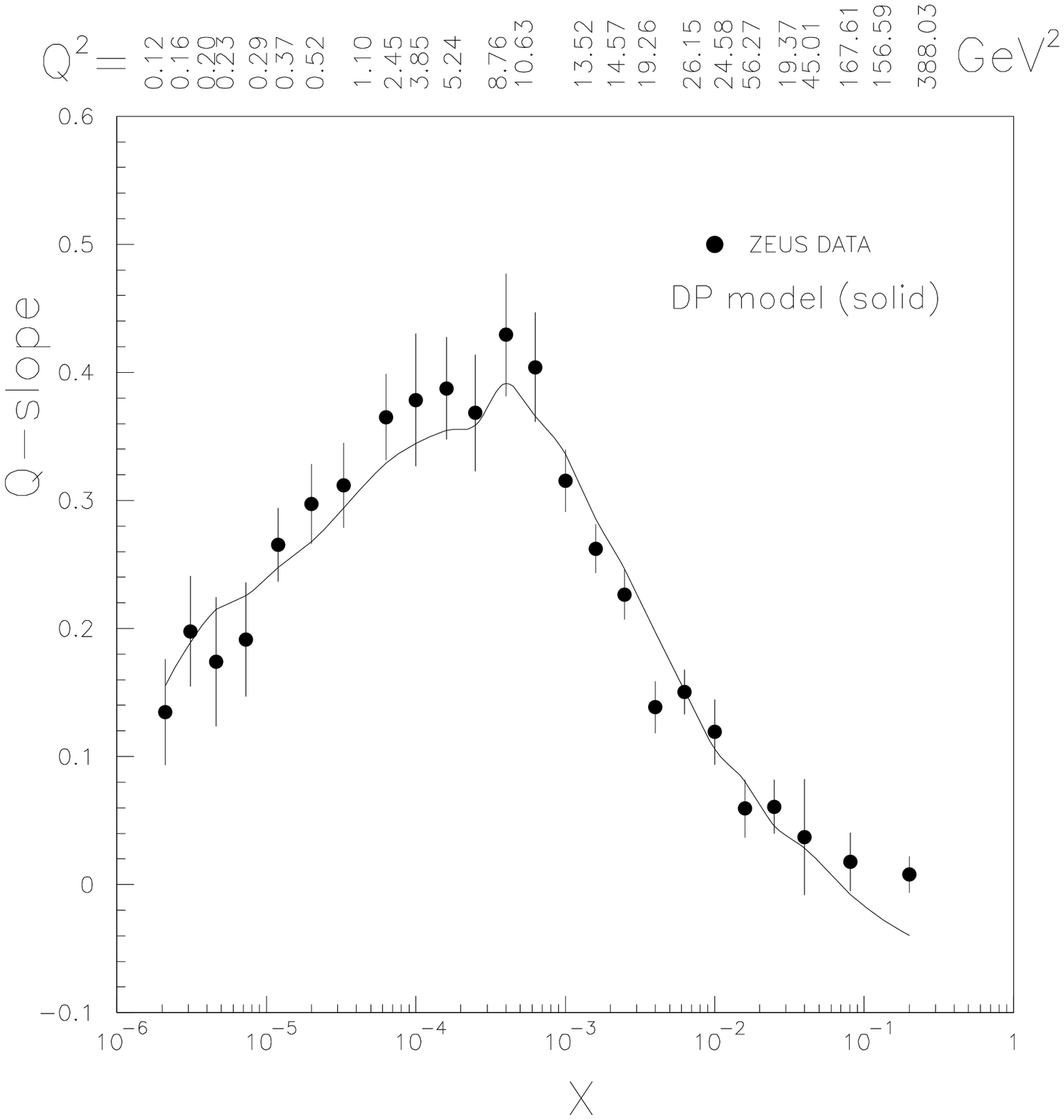,width=12cm}
\end{center}
\bigskip
{\bf Fig.~3.}
Q-slope
$B_Q(x,<Q^2>)$~: experimental points from  \cite{ZEUS98} as 
a function of $x$ (for the indicated $<Q^2>$ values). The 
continuous line is the prediction for the
Q-slope ${dF_2^p\over d\ell n Q^2}$ calculated (not fitted) within the Dipole 
Pomeron model.

\begin{center}
\epsfig{figure=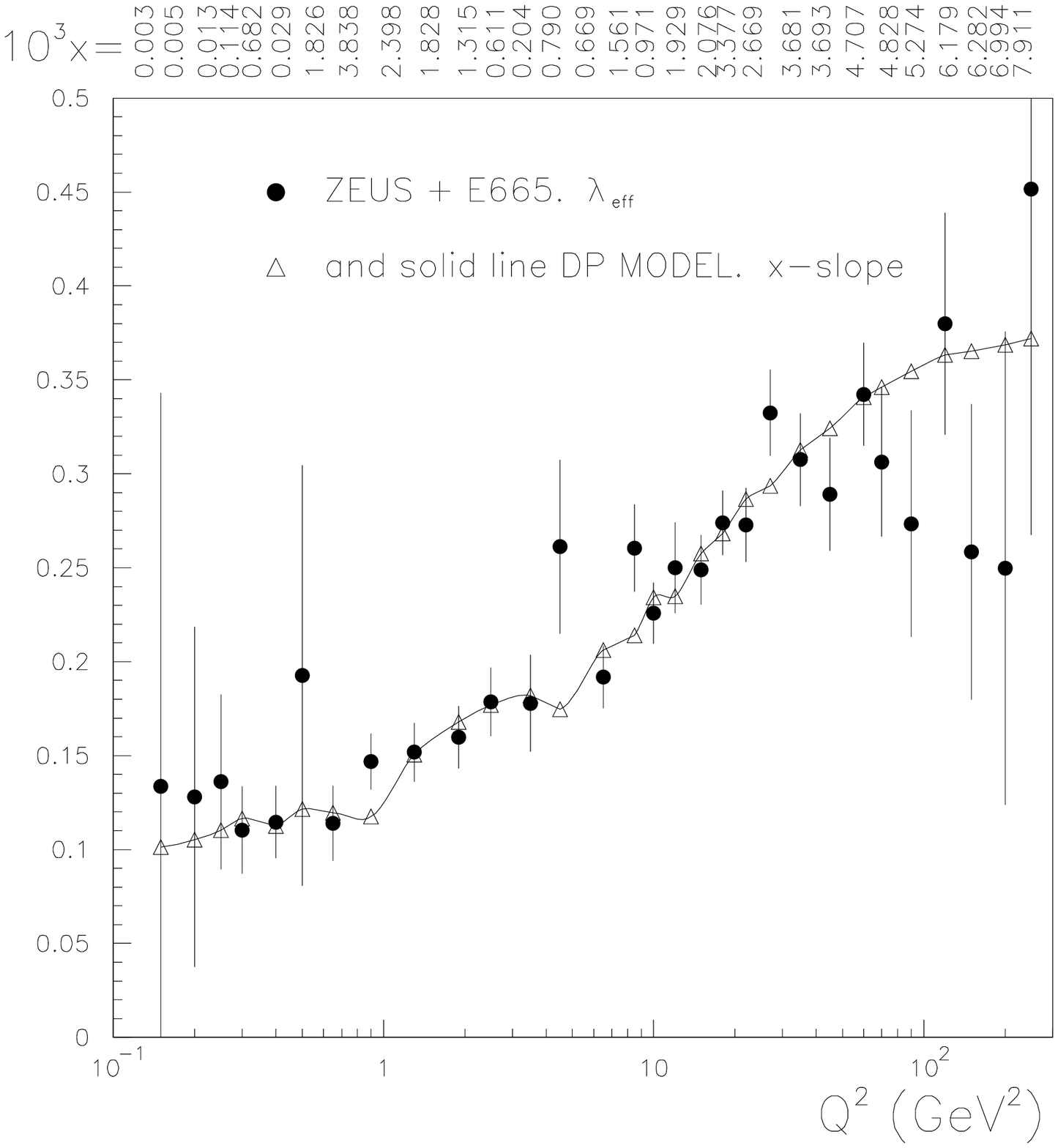,width=12cm}
\end{center}
\bigskip
{\bf Fig.~4.}
Experimental effective power $\lambda_{eff}(<x>,Q^2)$~: data from~\cite{ZEUS98} as 
a function of $Q^2$ (for the indicated $<x>$ values). The continuous line is the
predictions for the 
$x$-slope ${\partial \ell n F_2^p\over\partial\ell n (1/x)}$ calculated
(not fitted) in the Dipole Pomeron model.

\end{document}